\documentclass[twocolumn,showpacs,preprintnumbers,amsmath,amssymb]{revtex4}
\usepackage{graphicx}
\usepackage{dcolumn}
\usepackage{bm}

\begin{document}

\title{Noise-Induced Phase Space Transport in Time-Periodic Hamiltonian Systems}

\author{Bal{\v s}a Terzi{\'c}}
  \email{bterzic@astro.ufl.edu}
  \affiliation{Department of Astronomy, University of Florida, Gainesville, 
               Florida 32611}

\author{Henry Kandrup}
  \altaffiliation{Passed away on October 18, 2003}
  \affiliation{Department of Astronomy, Department of Physics, and 
               Institute for Fundamental Theory \\
               University of Florida, Gainesville, Florida 32611}

\date{\today}

\begin{abstract}
Orbits in a three-dimensional potential subjected to periodic driving,
$V(x^{i},t)=(1+m_{0}\sin {\omega}t)V_{0}(x^{i})$, divide naturally into two 
types, regular and chaotic, between which transitions are seemingly impossible.
The chaotic orbits divide in turn into two types, apparently separated by
entropy barriers, namely `sticky' orbit segments, which are `locked' to the
driving frequency and exhibit little systematic energy diffusion, and `wildly'
chaotic segments, which are not so locked and can exhibit significant energy
diffusion. Attention focuses on how the relative abundance of these different
orbit types and the transition rate between sticky and wildly chaotic orbits
depends on amplitude $m_{0}$, and on the extent to which these quantities can
be altered by weak friction and/or noise and by pseudo-random
variations in the driving frequency, idealized as an Ornstein-Uhlenbeck
process with (in general) nonzero autocorrelation time $t_{c}$.
When, in the absence of perturbations, there exist large measures of both
regular and chaotic orbits, the primary effect of weak noise is to increase
the relative measure of chaotic orbits. Alternatively, when almost all the
orbits are already chaotic, noise serves primarily to accelerate transitions
from sticky to wildly chaotic behavior. The presence or absence of friction
is unimportant and the details of the noise seem largely immaterial. In
particular, there is only a weak, roughly logarithmic dependence on amplitude,
and additive and multiplicative noise typically have virtually identical
effects. By contrast, allowing for high frequency `noisy' variations in
${\omega}$ tends to weaken effects of the driving, decreasing the relative
measure of chaotic orbits and suppressing large scale energy diffusion.
For both ordinary noise and noisy frequencies, the largest effects are 
observed for $t_{c}\to 0$, the efficacy of the noise decreasing with 
increasing $t_{c}$.

\end{abstract}

\maketitle

\section{INTRODUCTION AND OVERVIEW}
 \label{sec:level1}

Rapid energy diffusion in nearly periodic, oscillating potentials is an
important phenomenon arising in a variety of physical settings, including
collective relaxation in many-body systems interacting via long-range forces. 
One example is `violent relaxation'~\cite{bertin,LB}, the process whereby a 
galaxy involved in a collision or close encounter with another galaxy evolves 
towards a (quasi-) equilibrium. Another is halo formation in an accelerator 
beam~\cite{reiser}, where particles in an initially concentrated bunch are 
ejected to relatively large distances from the center.

This efficient energy diffusion can often be attributed to parametric
resonances between the frequency of the periodic driving and the natural
frequencies of the orbits ({\em e.g.,} \cite{gluckstern}) which, for
appropriate driving frequencies, tend to make many otherwise regular orbits
strongly chaotic~\cite{kvs}. However, not all chaotic orbits exhibit
systematic energy diffusion.  In at least some time-dependent spherically 
symmetric systems chaotic orbit segments divide empirically into two 
types~\cite{tk}.

On the one hand, there are `sticky' chaotic segments which are `locked' to
harmonics of the driving frequency and hence, albeit chaotic with positive
finite time Lyapunov exponents, do not exhibit systematic drifts in energy.
On the other, there are `wildly' chaotic segments which are not so locked and,
as such, can and do drift in such a fashion as to exhibit large-scale energy
diffusion.  However, this distinction between `sticky' and `wildly' chaotic is
not absolute. If, {\em e.g.,} a sticky orbit is integrated long enough, it can
become unstuck and start behaving as a wildly chaotic orbit. Nevertheless,
the time required for such a transition can be quite long, hundreds of
orbital times $t_{D}$ or more.

This transitional behavior appears analogous, at least qualitatively, to
diffusion through cantori in two-degree-of-freedom Hamiltonian system.
Cantori act as `entropy barriers' which impede, but do not prevent, phase
space transport~\cite{mmp}. However, barrier penetration in such systems can
be dramatically accelerated by allowing for noise~\cite{lw}\cite{pk99} which,
by wiggling the orbits, helps them to `find' suitable holes.

In this setting noise too acts via a resonant coupling \cite{pk99}. White
noise, which has zero autocorrelation time and power at all frequencies,
tends to be comparatively efficient as a source of accelerated phase space
transport.  By contrast, colored noise, with a finite autocorrelation time,
tends to be efficient only if the autocorrelation time is less than or
comparable to the orbital time scale, so that the noise has substantial power
at frequencies for which the orbits have power.

An obvious question, therefore, is whether transitions between sticky and
wildly chaotic orbits in time-dependent Hamiltonian systems can also be
accelerated by noisy perturbations. Here there are (at least) two different
types of `noise' to consider. There is of course the possibility of
`ordinary' noise, either intrinsic or extrinsic (see, {\em e.g.,} \cite{vanK}),
incorporated simply as an extra term in the equations of motion, just as for
a time-independent Hamiltonian system. However, there is also the possibility
of allowing for a `noisy' driving frequency, {\em i.e.,} allowing the
frequency to vary in a (near-)random fashion. Even a modest `wiggling' of
the frequency might weaken the resonant couplings, thus decreasing the effects
of periodic driving.

In all this, one is interested primarily in the effects of low-amplitude
perturbations. After all, it is obvious that, for sufficiently large
perturbations, solutions to the perturbed problem will be very different
from solutions to the unperturbed problem. The real question is {\it whether
relatively small perturbations, which  act in `real' systems but which
one might naively exclude in physical modeling, are sufficiently important
that they cannot be safely ignored}.
For example, even relatively low amplitude oscillations in
the bulk potential associated with a charge-particle beam can prove important
by facilitating the ejection of charges to an extended outer halo; and
allowing for modest frequency variations can actually make the effect more
pronounced~\cite{BS}.

Section II focuses on the empirical distinctions between sticky and wildly
chaotic orbits in a strictly periodic potential, addressing, in particular,
the issue of robustness. Earlier work established that these distinctions
exist in spherically symmetric potentials. However, the situation is less 
clear for more complex potentials. What happens if, for instance, chaotic 
orbits exist even in the absence of the periodic driving?  Could this tend 
to decrease the relative measure of sticky orbit segments or, perhaps, 
eliminate them altogether?  Section III focuses on the role of 
`ordinary' noise, modeled as an Ornstein-Uhlenbeck process; and Section 
IV considers the role of random variations in the driving frequency.
Section V summarizes the principal conclusions and then comments on potential
implications.

\section{CHAOS AND ENERGY DIFFUSION}
The objective here is to show that absolute distinctions between regular 
and chaotic orbits and short-time distinctions between `sticky' and `wildly' 
chaotic orbit segments, observed~\cite{tk} in time-dependent one-dimensional, 
{\em e.g.,} spherically symmetric, potentials persist in more realistic 
time-dependent three-dimensional potentials; and to investigate for several 
examples how the relative measures of different orbit types depend on 
properties of the potential.

Attention focuses on time-dependences of the form
\begin{equation}
V(x^{i},t)=(1+m_{0}\sin{\omega}_{0}t)\times V_{0}(x^{i}),
\end{equation}
with $V_{0}$ an unperturbed time-independent potential.
Two classes of potentials $V_{0}$ were considered, namely
\begin{equation}
V_{0}(s,t)=-{1\over (1+s^{2})^{1/2}}
\end{equation}
and
\begin{equation}
V_{0}(s,t)=-{1\over 1+s},
\end{equation}
each with
\begin{equation}
s^{2}=(x/a)^{2}+(y/b)^{2}+(z/c)^{2}.
\end{equation}
For spherical systems, with $a=b=c=1$, these reduce to integrable potentials 
well known from galactic astronomy. The former corresponds via the Poisson 
equation to a so-called Plummer density distribution with a smooth central 
core (see, {\em e.g.,} \cite{bertin}); the latter yields a ${\gamma}=1$ 
cuspy Dehnen~\cite{deh} density distribution similar to that observed in many
galaxies~\cite{lauer}.  The nonspherical generalizations of the Plummer 
potential tend not to admit many chaotic orbits but, for appropriate choices 
of axis ratios, the nonspherical Dehnen potentials do admit significant 
measures of chaotic orbits.  For the intermediate energies 
$E{\;}{\sim}{\;}-0.8$ used for the numerical experiments described in this 
paper, a typical orbital time scale $t_{D}{\;}{\sim}{\;}5$.

Ensembles of ${\ge}{\;}800$ initial conditions were generated by uniformly
sampling an (unperturbed) constant energy hypersurface, and these initial
conditions were then evolved for a time $t{\;}{\ge}{\;}4096$, allowing for
a driving frequency ${\omega}_{0}$ so chosen as to trigger parametric
resonance. The integrations also solved simultaneously for the evolution of
a linearized perturbation, thus facilitating an estimate of the largest finite 
time Lyapunov exponent ${\chi}$.  Orbital data recorded at fixed intervals 
were Fourier-transformed and the resulting data analyzed to establish 
correlations between the value of the largest Lyapunov exponent and the 
frequencies for which the $x$-, $y$-, and $z$-coordinates of the orbits have 
the most power.

For all the models that were tested, involving pulsations of both integrable 
and nonintegrable potentials $V_{0}$, the orbits divide naturally into two 
distinct types, namely regular and chaotic, between which transitions appear 
impossible. Estimates of the largest Lyapunov exponent for the regular orbits 
decay towards zero in the usual way (see, {\em e.g.,}~\cite{bgs});
for chaotic orbits they remain strictly positive~\cite{inf}.
Moreover, regular orbits appear multiply periodic, whereas the Fourier spectra
of chaotic orbits are broader band.

It also appears that chaotic orbit segments subdivide into `sticky' and
`wildly' chaotic.  Sticky chaotic segments are clearly chaotic in the sense
that they have positive finite time Lyapunov exponents and more complex
Fourier spectra than do regular orbits. However, the spectra do exhibit
striking regularities. In particular, the frequencies for which the transforms
of the spatial coordinates have the most power are almost exactly equal to
a harmonic of the driving frequency ${\omega}_{0}$.  In many cases almost 
all the power is at that special frequency; in others there is substantial 
power at nearby, mostly lower, frequencies but the driving frequency is 
still clearly dominant. That the orbits appear `locked' to a harmonic might 
suggest that their energies cannot change significantly, a prediction which 
is readily confirmed. Plots of $E(t)$ for sticky orbit segments often look 
nearly periodic visually, and the associated Fourier spectra $|E({\Omega})|$ 
tend to be very sharply peaked (although there is always some `scruff' 
indicating that the energy is not {\em exactly} periodic).  Wildly chaotic 
orbits have substantially more complex spectra, and they are not obviously 
locked to any harmonic of the driving frequency.  Rather, as viewed in 
configuration or energy space, or as characterized by their evolving spectral 
distributions, it is evident that these wildly chaotic orbits can, and 
often do, diffuse more or less freely through phase space.

Unlike the distinction between regularity and chaos, the distinction between
sticky and wildly chaotic is not absolute. If an initially sticky chaotic
orbit segment is evolved for a sufficiently long period, it will eventually
transform itself into a wildly chaotic orbit. As discussed elsewhere~\cite{tk},
this suggests that, in the presence of the driving, the chaotic phase space
regions are partitioned by `entropy barriers', corresponding, {\em e.g.,} to
small holes through which chaotic orbits eventually effuse.

Examples of these different types of orbits are exhibited in Fig.~1. Here
the top row corresponds to a regular orbit, the middle two rows to sticky
chaotic segments, and the bottom two to wildly chaotic segments. One of
these wildly chaotic segments remains bound energetically ($E<0$) for the 
duration of the integration. The other becomes unbound energetically before 
$t=2048$. In each case, the leftmost column exhibits the trajectory of the 
orbit as viewed in terms of the phase space coordinates $x$ and $v_{x}$, 
each recorded at intervals ${\delta}t=1.0$. The second column exhibits the 
mean drift in energy, ${\delta}E/E$, as a function of time, and the third 
exhibits the Fourier transform of the resulting time series. The fourth 
column exhibits estimates of the largest Lyapunov exponent ${\chi}$, 
computed for times $<t$, and the last column exhibits the power spectrum 
$|x({\Omega})|$.
\begin{figure*}
\includegraphics{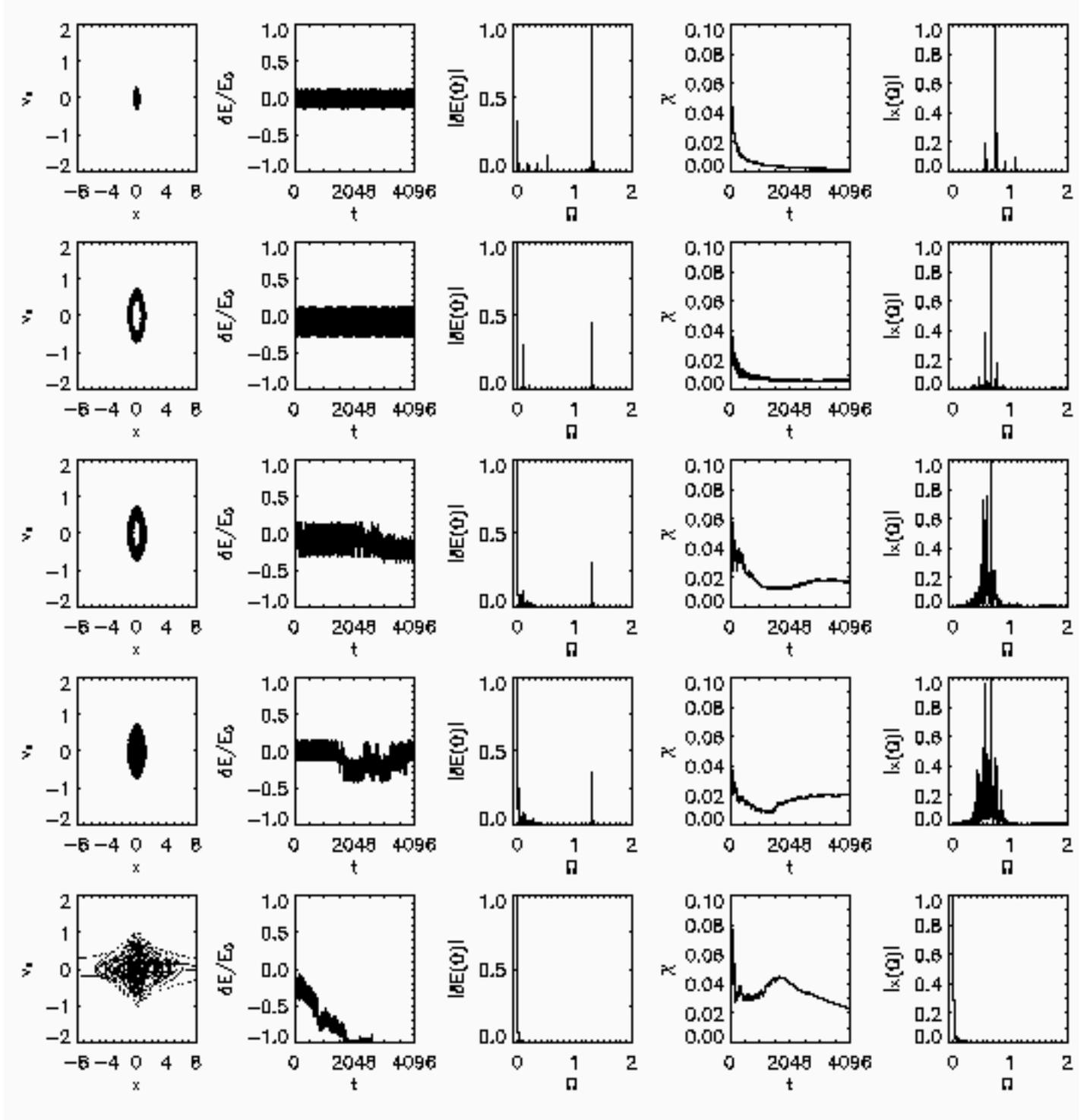}
\caption{
Top row: the phase space trajectory of a typical regular orbit evolved in a 
triaxial Plummer potential with axis ratios $a^{2}:b^{2}:c^{2}=1.25:1:0.75$, 
with initial energy $E=-0.8$, $m_0=0.1$ and $\omega_{0}=1.3$, along with the 
relative energy oscillations, $\delta E(t)/E$, the power spectrum 
$\left|\delta E(\Omega) \right|$, the finite time Lyapunov exponents $\chi$ 
computed for times $T \le t$, and the power spectrum, $\left|x(\Omega)\right|$.
Second row: the same for a `nearly regular' sticky chaotic orbit, with the 
peak frequency for $|x({\Omega})|$ satisfying ${\Omega}^{(1)}={\omega}_{0}$ 
and the peak frequency for $|E({\omega})|$ equally ${\omega}_{0}/2$. 
Third row: a more irregular sticky orbit which still has the same peak 
frequencies. Fourth row: a wildly chaotic orbit with spectra peaked at lower
frequencies. Fifth row: a more wildly chaotic orbit that rapidly diffuses 
toward larger radii and lower frequencies.
}
\end{figure*}

As noted already, the type of orbit correlates not only with the overall
shape of the Fourier spectrum, but with the value of the peak frequencies
for which quantities like $|x({\Omega})|$ or $E({\omega})$ have the most 
power. This is illustrated in Fig.~2, which exhibits scatter plots of 
${\chi}$, the largest finite time Lyapunov exponent, and ${\Omega}^{(1)}$,
the frequency for which $|x({\Omega})|$ has the most power. Each panel was
generated from the same set of $800$ initial conditions. The four columns
from left to right correspond to increasing amplitudes $m_{0}=0$,
$m_{0}=0.05$, $m_{0}=0.1$, and $m_{0}=0.2$.  The driving frequency 
${\omega}_{0}=1.3$, and the orbital structure is dominated by a $2:1$ 
resonance. In particular, for sticky orbits $|x({\Omega})|$ peaks at 
${\Omega}^{(1)}={\omega}_{0}/2=0.65$. (By contrast, as is evident from 
Fig.~1, $|E({\Omega})|$ peaks at ${\Omega}^{(1)}={\omega}_{0}=1.3$.)
For all the regular orbits, ${\Omega}^{(1)}$ is larger than ${\omega}_{0}/2$,
the value of which is indicated by the vertical dashed line. Alternatively,
orbit segments with ${\Omega}^{(1)} \le {\omega}_{0}/2$ correspond to wildly 
chaotic orbits.
\begin{figure}
\includegraphics[width=3in]{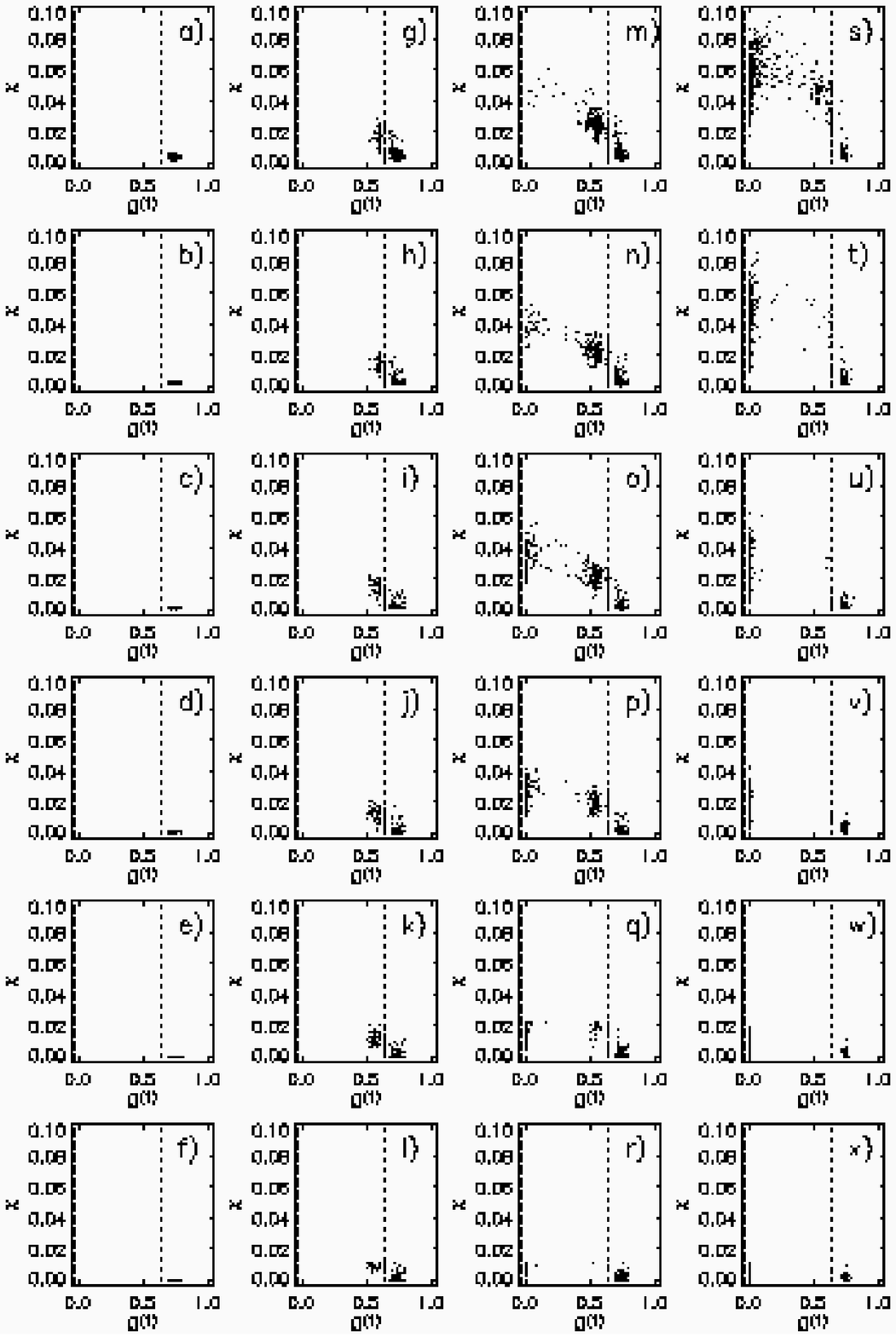}
\caption{
Finite time Lyapunov exponents as a function of peak frequency $\Omega^{(1)}$
for regular ($\Omega^{(1)} > \omega_{0}/2$), sticky ($\Omega^{(1)} \approx 
\omega_{0}/2$), and wildly chaotic ($\Omega^{(1)} < \omega_{0}/2$) orbits, 
generated from a microcanonical sampling of 800 initial conditions with 
energy $E=-0.8$, evolved in a triaxial Plummer potential with 
$a^{2}:b^{2}:c^{2}=1.25:1:0.75$ and driving frequency $\omega=1.3$, integrated 
for times $t=1024$, $2048$, $4096$, $8192,$ and $16384$, with amplitudes: 
(a)-(f) $m_0=0$, (g)-(l) $m_0=0.05$, (m)-(r) $m_0=0.1$, (s)-(x) $m_0=0.2$.
}
\end{figure}

For lower amplitude driving, the relative measure of chaotic -- as opposed
to regular -- orbits and the typical size of the largest finite time Lyapunov
exponents for the chaotic orbits both tend to be smaller. In particular, for
energies and potentials which admit no chaos in the absence of periodic
driving, there appears to exist a minimum threshold amplitude,
$m_{0}{\;}{\sim}{\;}0.05$, for the onset of appreciable amounts of chaos.
For lower amplitudes, the relative measure of sticky chaotic segments --
as opposed to wildly chaotic segments -- also tends to be higher and, as 
will be discussed later, in some cases the time required for transitions 
from sticky to wildly chaotic tends to be longer.

Of especial interest is the case of potentials which, in the absence of
periodic driving, admit large measures of both regular and chaotic orbits.
As illustrated in Fig.~3, in this case weak driving tends to make some, albeit
typically not all, the otherwise regular orbits chaotic; and the already
chaotic orbits tend to remain chaotic. Few if any otherwise chaotic orbits
are converted to regular. However, it is not true that the driving
necessarily increases the values of the largest finite time Lyapunov exponents
for all the otherwise chaotic orbits. Indeed, for sufficiently small
$m_{0}$ the driving is comparably likely to increase or decrease the value
of ${\chi}$.  Significantly, though, as $m_{0}$ is increased the degree of
chaos always appears to increase, at least initially. For sufficiently large
values of $m_{0}$ all the otherwise regular orbits become chaotic and, for the
otherwise chaotic orbits, there is a  systematic increase in the values of
the largest ${\chi}$. However, there will still be a few otherwise chaotic
orbits which become less chaotic in the presence of driving.  One other 
obvious trend evident from Fig.~3 is the systematic decrease in the values 
of ${\chi}(m_{0}{\ne}0)$ at late times. This decrease reflects the fact 
that the vast majority of the chaotic orbits have become wildly chaotic, 
and that many of these have drifted to less negative energies, where the 
orbital time $t_{D}$, and hence the Lyapunov time $1/{\chi}$, is very long.
\begin{figure}
\includegraphics[width=3in]{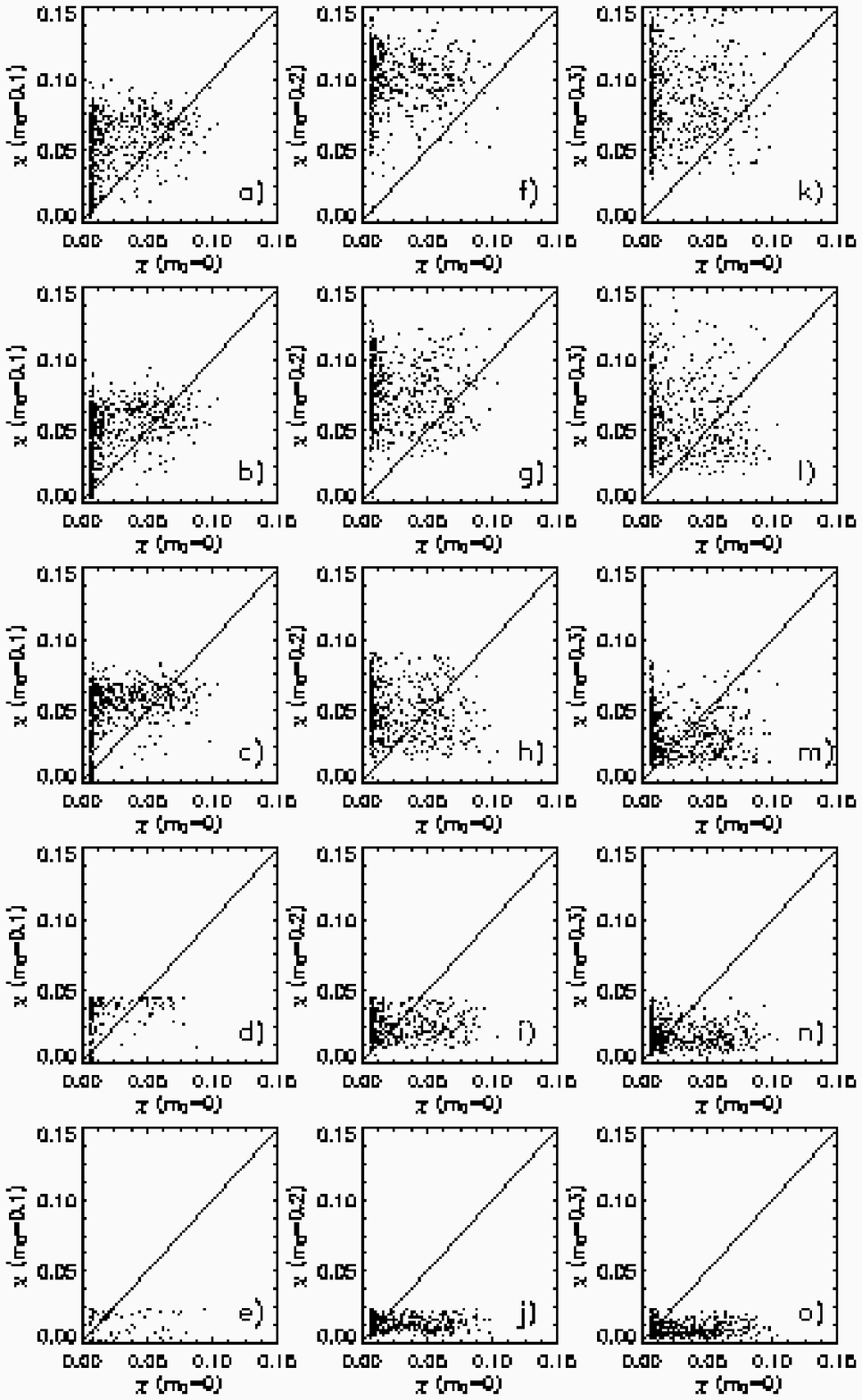}
\caption{
Finite time Lyapunov exponents for driven versus unperturbed orbits, integrated
from the same microcanonical set of $800$ initial conditions with $E=-0.8$
in a triaxial Dehnen potential with axis ratios $a^{2}:b^{2}:c^{2}$
$=1.25:1:0.75$ and driving frequency $\omega=2.75$ at varying integration
times $t=1024$, $2048$, $4096$, $8192$, and $16384$ for: (a) - (e) $m_0=0.1$,
(f) - (j) $m_0=0.2$ and (k) - (o) $m_0=0.3$.
}
\end{figure}

Whether, in the presence of periodic driving, an initial condition results
initially in sticky as opposed to wildly chaotic behavior does not appear to
correlate strongly with its behavior in the absence of driving. For example,
there is no obvious preferential tendency for otherwise regular initial
conditions to yield sticky behavior. For a fixed potential, the likelihood 
that an initial condition yields a sticky, as opposed to wildly chaotic, 
orbit is determined more by the amplitude $m_{0}$ than by the regular or 
chaotic behavior exhibited by the initial condition in the absence of 
periodic driving. When $m_{0}$ is small, sticky orbits tend to be more 
common; as $m_{0}$ increases, the relative measure of wildly chaotic segments 
increases.

The relative abundance of different orbit types does not appear to exhibit
a simple, universal dependence on shape. In some cases, there is more chaos
for parameter values where the equipotential surfaces are oblate and/or
prolate, rather than spherical; in others spherical potentials exhibit more 
chaos. The one trend that {\em does} emerge, at least for the models 
considered here, is that genuinely triaxial potentials with 
$a{\;}{\ne}{\;}b{\;}{\ne}{\;}c$ tend to exhibit more chaos than potentials 
with either spherical or spheroidal symmetry.

To quantify at least partially the transition rate between sticky and wildly
chaotic behavior, one can, for an orbit ensemble of fixed initial energy,
begin by identifying the initial conditions that correspond to chaotic orbits,
then identifying the range of energies which, for these initial conditions,
correspond to sticky behavior, and finally determining for each orbit
generated from a chaotic initial condition the first time that it acquires 
an energy outside this range. The obvious quantity of interest is the 
fraction $N(t)/N(0)$ of those chaotic orbits which have not yet `escaped' 
within time $t$. When, as for the example presented below, the orbital 
structure is dominated by a $2:1$ resonance, regular orbits can be rejected 
immediately as having ${\Omega}^{(1)}>{\omega}_{0}/2$.  However, the computed 
fraction does depend somewhat on the precise choice of energy range, which 
is not completely obvious. Fortunately, though, as discussed below, modest 
changes do not significantly alter the final results.

Examples of such an analysis, all generated for the same ensemble, are
exhibited in Fig.~4, each panel of which plots $N(t)/N(0)$, the relative number
of chaotic orbits which are sticky, as a function of time $t$ for different
choices of amplitude $m_{0}$. The top two panels are for oblate
($a^2=b^2=1$, $c^2=0.75$) and triaxial ($a^2=1.25$, $b^2=1$, $c^2=0.75$)
Plummer potentials which, in the absence of driving, admit few if any chaotic
orbits.  The lower panels are for oblate and triaxial Dehnen potentials with
the same axis ratios, which admit a considerable amount of chaos even in the
absence of driving. From this Figure two points are immediately evident. The
first is that, in every case, after a very rapid initial decrease,
$\ln N$ decreases systematically in a fashion that is roughly linear in time.
This implies that, at least in a first approximation, transitions from sticky
to wildly chaotic can be viewed a Poisson process,
\begin{equation}
N(t)=N_{0}\exp(-{\Lambda}t),
\end{equation}
with $N_{0}$ the number of chaotic orbits which are initially sticky
and ${\Lambda}$ the escape rate.  This exponential decrease is the same 
behavior observed for diffusion through cantori in two-degree-of-freedom 
systems (see, {\em e.g.,} \cite{pk99}), thus reinforcing again the notion 
that these transitions involve passage through an entropy barrier.
\begin{figure}
\includegraphics[width=3in]{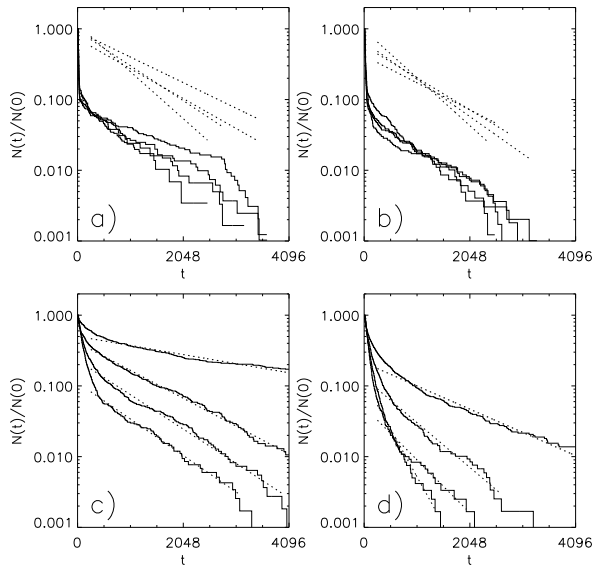}
\caption{
(a) Fraction $N(t)/N(0)$ of chaotic orbits that remain `sticky' after time $t$,
computed for ensembles of $1200$ initial conditions with energy $E=-0.8$, 
evolved in an oblate Plummer potential with axis ratio 
$a^{2}:b^{2}:c^{2}=1:1:0.75$, pulsed with frequency ${\omega}_{0}=1.3$
and variable amplitude $m_{0}$.  Solid curves with increasing thickness
correspond to increasing amplitudes $m_{0}=0.15,0.2,0.25,0.3$.  Dashed lines
represent best linear fits.  (b) The same for a triaxial Plummer potential with
$a^{2}:b^{2}:c^{2}=1.25:1:0.75$, again with $E=-0.8$ and ${\omega}_{0}=1.3$.
(c) The same for an oblate Dehnen potential with $a^{2}:b^{2}:c^{2}=1:1:0.75$, 
again with $E=-0.8$ but now with ${\omega}_{0}=2.75$.  (d) The same for a 
triaxial Dehnen potential with $a^{2}:b^{2}:c^{2}=1.25:1:0.75$ with 
$E=-0.8$ and ${\omega}_{0}=2.75$.
}
\end{figure}

The other striking point is that the dependence of ${\Lambda}$ on $m_{0}$
is very different for the Plummer and Dehnen models. For the Plummer models 
${\Lambda}$ is rather nearly independent of driving amplitude; but, for 
the Dehnen models ${\Lambda}$ increases significantly as $m_{0}$ increases. 
This is an example of a more general result: If, in the absence of driving, 
the orbits are all regular, ${\Lambda}$ exhibits little if any dependence 
on $m_{0}$. If, however, a significant fraction of the orbits are already 
chaotic in the absence of driving, ${\Lambda}$ is a strongly increasing 
function of $m_{0}$. The degree of chaos exhibited by a model, as probed 
by the relative measure of chaotic orbits or the size of a typical finite 
time Lyapunov exponent, need not correlate especially with the amount of 
chaos that is present in the absence of driving. However, the presence or 
absence of chaos in the unperturbed models {\em does} impact the transition 
rate between sticky and wildly chaotic behavior.

But why do plots of $N(t)$ show an initial much more rapid decrease, and how
sensitive are these results to the energy range associated with sticky orbit
segments? The basic conclusion is that increasing the size of the sticky
range slows the early time rapid decrease, but has only a minimal impact on
the computed transition rate ${\Lambda}$. The results exhibited in Fig.~4
were computed assuming a range $E_{0}-1.5m_{0}<E<E_{0}+1.5m_{0}$, with $E_0$
the initial unperturbed energy and $m_{0}$ the amplitude of the perturbation,
but, in most cases, the best fit ${\Lambda}$ changes only by $10\%$ or so if
$1.5$ is increased to $3.0$~\cite{weird}. The validity of this criterion was
also tested by comparing the number of orbits that this prescription classifies
as sticky at time $t=4096$ with the number of orbits which still appear
locked to the driving frequency with ${\Omega}^{(1)}={\omega}_{0}/2$.

The obvious interpretation is that the initial rapid decrease
reflects wildly chaotic orbits which require a finite time to escape from
the sticky range: if the range is made larger, the time required for escape
becomes longer. This suggests in turn that the most accurate estimate of the
number of sticky orbits involves assuming the smallest range which clearly
includes all sticky orbits, the criterion used to justify a value
${\approx}{\;}1.5$.  Given such a choice, one also appears justified in
interpreting $N_{0}$ in eq.~(2.5) as the relative fraction of the chaotic
orbits which are sticky at or near $t=0$.

Given this interpretation of $N_{0}$, one can easily estimate the relative
abundance of regular, sticky, and wildly chaotic orbits at time $t=0$. The
results for the ensembles used to construct Fig.~4 are exhibited in Fig.~5.
It is evident that, in each case, the relative number of chaotic orbits is
an increasing function of $m_{0}$, but the relative number of initially
sticky orbits can vary significantly. For the Plummer potentials, the relative
measure of sticky orbits tends to be very small irrespective of the magnitude
of $m_{0}$. For the Dehnen potentials, however, sticky orbits are comparatively
common, especially for smaller amplitude driving. In particular, for
$m_{0}{\;}{\le}{\;}0.15$ there are more sticky than wildly chaotic orbits for
the oblate Dehnen potential.  It should, however, be stressed that the
relative measure of sticky versus wildly chaotic model need not increase
monotonically as a potential is made less symmetric. For example, ensembles
evolved in a spherical Plummer potential perturbed with the same amplitude
$m_{0}$ tend to exhibit substantially larger relative measures of sticky orbits.
\begin{figure}
\includegraphics[width=3in]{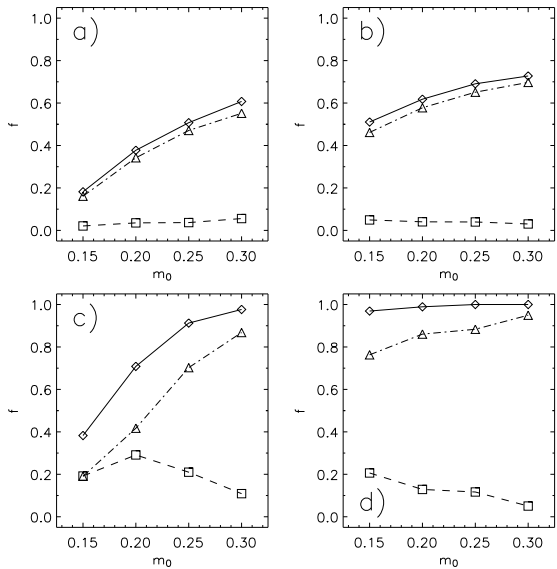}
\caption{
(a) Relative number of chaotic orbits (solid line and diamonds), sticky 
chaotic orbits (dashes and squares), and wildly chaotic orbits (dot-dashed
lines and triangles) at $t=0$, plotted as a function of $m_{0}$, for the
ensembles in Fig.~4, as predicted by the best linear fits.
}
\end{figure}

Assuming that these conclusions are robust, they have obvious implications 
for chaotic phase mixing in systems subjected to periodic perturbations. 
{\it Relatively low amplitude periodic driving can suffice to trigger large 
measures of chaotic orbits}.  However, these chaotic orbits are apt to be 
primarily sticky and, as such, unlikely to exhibit large-scale secular 
changes in energy. In the context of charged-particle beams, this would 
suggest that periodic driving could facilitate a comparatively rapid loss 
of coherence, but that it would not necessarily trigger the formation of a 
halo. If, however, the amplitude of the driving is increased, a larger 
measure of the orbits will be wildly chaotic; and a substantial fraction 
of these could drift to higher energy states associated with an extended 
halo. This expectation is consistent with real experiments involving 
charged-particle beams ({\em e.g.,}~\cite{haber}), which indicate that 
particle bunches which undergo larger amplitude oscillations (larger initial 
`mismatch') tend to form more extended halos.

\section{ LOW AMPLITUDE NOISE}
The objective now is to describe how distinctions amongst regular, sticky 
chaotic, and wildly chaotic behavior and, especially, the rate of transitions 
between sticky and wildly chaotic behavior, are impacted by the introduction 
of weak perturbations, modeled as friction and/or Gaussian noise. The 
conclusions derive from an analysis of orbits generated as solutions to the 
perturbed evolution equation
\begin{equation} \label{langevin}
{d^{2}x^{i}\over dt^{2}}=-{{\partial}V\over {\partial}x^{i}}
-{\eta}{dx^{i}\over dt}+F^{i}(t).
\end{equation}
Here ${\eta}dx^{i}/dt$ represents a dynamical friction and $F^{i}$ is a
Gaussian-distributed random force which, for specificity, is assumed to sample
an Ornstein-Uhlenbeck process.

In the context of a galaxy or charged-particle beam, random perturbations
associated with an external environment could correspond~\cite{vanK} to
extrinsic noise, for which ${\eta}{\;}{\equiv}{\;}0$. Alternatively, effects
associated with graininess or internal substructures could correspond to
intrinsic noise, with a nonzero ${\eta}$ related to the moments of $F^{i}$ by
the Fluctuation-Dissipation Theorem. In particular, the effects of graininess
can be modeled (see, {\em e.g.,}~\cite{sk02}) as intrinsic white noise, which
can be viewed as a singular limit of an Ornstein-Uhlenbeck process with
vanishing autocorrelation time.

The assumption that $F^{i}$ is Gaussian noise sampling an Ornstein-Uhlenbeck 
process implies that the stochastic process is characterized completely by 
its first two moments,
\begin{displaymath}
{\langle}F^{i}(t){\rangle}=0 \qquad {\rm and}
\end{displaymath}
\begin{equation} \label{Fc}
{\langle}F^{i}(t_{1})F^{j}(t_{2}){\rangle}= {\delta}^{ij}\;K(|t_{1}-t_{2}|),
\end{equation}
where ${\langle}\;.\; {\rangle}$ denotes a statistical average and the 
autocorrelation function $K$ satisfies
\begin{equation}\label{Kc} 
K(|t_{1}-t_{2}|) = {\Delta}^{2}\exp(-|t_{1}-t_{2}|/t_{c}).
\end{equation}
Here ${\Delta}$ represents the typical `size' of the random force and the 
autocorrelation time $t_{c}$ the time scale over which it changes appreciably. 
The limit $t_{c}\to 0$ corresponds to white noise.  State-independent, 
additive noise has ${\eta}$ and $K$ independent of phase space coordinates. 
Multiplicative noise arises when these quantities are functions of 
$x^{i}$ and/or $v^{i}$.

Three different variants were considered:
\par\noindent
1. Additive noise, both white and colored, without friction.
\par\noindent
2. Additive noise with friction related by the Fluctua-tion-Dissipation
Theorem,
\begin{equation} \label{FDT}
\int_{-\infty}^{+\infty}d{\tau}K({\tau})=2{\Theta}{\eta}{\;}{\equiv}{\;}D,
\end{equation}
where ${\Theta}$ represents a temperature and $D$ the diffusion constant
entering into a Fokker-Plank description~\cite{vanK}. (This implies
${\Delta}^{2}={\Theta}{\eta}/t_{c}$.) On physical grounds, ${\Theta}$ was 
typically selected to be comparable in magnitude to the initial energy of 
the orbit(s) in question.
\par\noindent
3. Multiplicative noise with $K(v^{i})$ satisfying
\begin{equation}
K(v^{i})=K_{0}v^{2}/{\langle}v^{2}{\rangle},
\end{equation}
with $K_{0}$ the autocorrelation function appropriate for additive noise,
$v^{2}$ the squared speed of the orbit, and ${\langle}v^{2}{\rangle}$
the mean squared speed of an orbit with the specified initial energy.

There is no uniformly accepted definition of chaos or Lyapunov exponent for
stochastic differential equations. However, the following prescription,
implemented in this paper, appears to capture one's physical intuition in
terms, {\em e.g.,} of the visual appearance of an orbit. Specifically, in the
context of this paper ${\chi}$ will be defined in the usual way by comparing
the evolution of two nearby initial conditions, {\em assuming}, however,
{\em that both orbits feel exactly the same noise and}, for consistency,
{\em the same friction}.  This entails solving a perturbation equation
\begin{equation}
{d^{2}{\delta}x^{i}\over dt^{2}}=
-{{\partial}^{2}V\over {\partial}x^{i}{\partial}x^{j}}{\delta}x^{j},
\end{equation}
%\par\noindent
where the Hessian is evaluated along the original noisy orbit. This
prescription obviates the necessity of deciding how to relate concrete
realizations of a stochastic process at two nearby phase space points.

Consider first the effects of white noise. If, in the absence of noise, most
of the orbits are not already chaotic, the most conspicuous effect resulting
from the addition of noise is to increase the relative abundance of chaotic
orbits. This is, for example, illustrated in Fig.~6, generated for the same
Plummer models used to produce Fig.~4, which exhibits the relative measure
$f_{c}$ of chaotic orbits as a function of amplitude ${\eta}$. For all three
choices of $m_{0}$, the noise has a discernible effect even for amplitudes as
small as ${\eta}=10^{-7}$; for ${\eta}$ as large as $10^{-4}$ almost all the
orbits have become chaotic. Interestingly, however, the noise does not have a
significant impact on the size of a typical finite time Lyapunov exponent
for those orbits which {\em are} chaotic.
\begin{figure}
\includegraphics[width=3in]{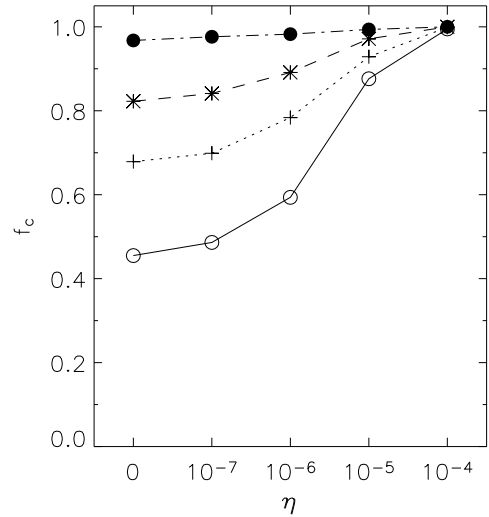}
\caption{
Relative measure of chaotic orbits, $f_{c}$, computed for a microcanonical
ensemble of $1200$ initial conditions with energy $E=-0.8$ in a pulsating 
triaxial Plummer potential with axis ratio $a^{2}:b^{2}:c^{2}=1.25:1:0.75$, 
evolved with $\omega_{0}=1.3$, $m_{0}=0.1$ (empty circles with solid line), 
$m_{0}=0.15$ (crosses with short-dashed line), $m_{0}=0.2$ (asterisks with 
long-dashed line) and $m_{0}=0.3$ (filled circles with dash-dotted line) in 
the presence of additive white noise with ${\Theta}=1.0$ and variable 
amplitude ${\eta}$.
}
\end{figure}

The empirical distinction between sticky and wildly chaotic orbits persists
in the presence of relatively low amplitude noise, and transitions between
sticky and wildly chaotic still appear to sample a (near-)Poisson process.
However, as was the case in the absence of noise, the observed behavior of
orbit ensembles depends considerably on whether or not the orbits are almost
all chaotic.

For systems like the Plummer potentials, where there is little chaos in the
absence of driving, the rate ${\Lambda}$ associated with transitions between
sticky and wildly chaotic is largely independent of ${\eta}$. Increasing the
noise amplitude, which increases the overall abundance of chaotic orbits, also
increases the abundance of initially sticky chaotic segments; but the rate
at which these are transformed into wildly chaotic orbits is largely unaffected
unless the noise becomes very large in amplitude.

Alternatively, for systems like the Dehnen potentials, where there is a good
deal of chaos even in the absence of driving, the addition of noise definitely
accelerates transitions from sticky to wildly chaotic. This is, {\em e.g.,}
illustrated in the left hand column of Fig.~7, which exhibits $N(t)/N(0)$,
the relative number of orbit that are still sticky, for the triaxial
Dehnen potential used to generate Fig.~4. The qualitative behavior manifested
in these panels is very similar to that associated with diffusion through
cantori in two-degree-of-freedom Hamiltonian systems ({\em cf.}~\cite{pk99}),
where one can argue that, by wiggling orbits, `noise helps orbits find holes'
in entropy barriers.
\begin{figure}
\includegraphics[width=3in]{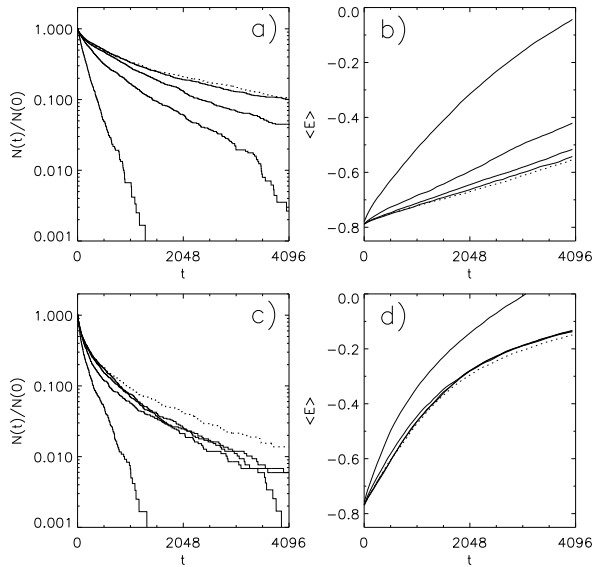} %fig7
\caption{
(a) Fraction $N(t)/N(0)$ of chaotic orbits that remain `sticky' after time $t$,
computed for ensembles of $1200$ initial conditions with $E=-0.8$, evolved in
a triaxial Dehnen potential with axis ratio $a^{2}:b^{2}:c^{2}=1.25:1:0.75$,
${\omega}_{0}=2.75$, and $m_0=0.1$, allowing for additive white noise with
variable amplitude. Dashed lines correspond to integrations without white
noise, while the magnitude of $\eta$ increases with increasing line thickness
through values $\eta=10^{-7}, 10^{-6}, 10^{-5}, 10^{-4}$. In each case
${\Theta}=1$. (b) The mean energy ${\langle}E(t){\rangle}$ computed for the
same ensemble.  (c) and (d) The same for $m_{0}=0.15$.
}
\end{figure}

The situation for larger amplitude noise is rather different. Seemingly
independent of the form of the potential, for ${\eta}>10^{-3}-10^{-2}$ escapes
happen so quickly that the distinction between sticky and wildly chaotic 
becomes essentially meaningless. This can be understood by observing that, 
when the noise has such a large amplitude, it will pump energy into the orbit 
so rapidly that, even if the driving has a relatively minimal effect, one 
cannot think of the orbit as being restricted to a nearly constant energy 
hypersurface.

Although these conclusions were derived  allowing only for additive
white noise in the absence of any dynamical friction, they appear to be
comparatively insensitive to the form of the perturbation. For example, the
presence or absence of a dynamical friction of comparable amplitude ${\eta}$
is largely irrelevant. Moreover, there is no indication that allowing for
`generic' multiplicative noise would alter the basic picture. For example,
allowing for multiplicative noise with an autocorrelation function scaling
as in eq.~(3.5) has only a minimal effect on the relative number of chaotic
orbits or the transition rate between sticky and wildly chaotic behavior.
All that really seems to matter is the amplitude of the noise. Moreover, even
this dependence on amplitude is comparatively weak. Inspection of Figs.~6 and
7 indicates that there is only a roughly logarithmic dependence on ${\eta}$.

Also of interest is how the combined effects of periodic driving and
friction/noise impact the mean energies of typical particle orbits.
Examples of such behavior, again computed for orbits in a pulsed, triaxial
Dehnen potential perturbed only by additive white noise, are exhibited in the
right hand column of Fig.~7.

But how are such data to be interpreted?
It follows immediately from the Langevin equation (3.1)
that, allowing only for periodic driving and additive
white noise, the mean energy ${\langle}E{\rangle}$ associated with some
ensemble will satisfy
\begin{equation} \label{meanE}
{d{\langle}E{\rangle}\over dt}=\left\langle
{{\partial}V\over {\partial}t}\right\rangle +3{\Theta}{\eta},
\end{equation}
where ${\langle}{\partial}V/{\partial}t{\rangle}$ represents an average
computed along the orbits. The noise leads directly to a linear increase
in energy at a constant rate $3{\Theta}{\eta}$, independent of the form of
the driving; the driving occasions an additional increase which will in general
depend on the form of the perturbed orbits. Given the mean value
${\langle}E(t){\rangle}$ computed for orbits that are subjected to both noise
and periodic driving, it is thus natural to subtract off the linear increase
$3{\Theta}{\eta}t$ directly attributable and to compare the residual with
the mean ${\langle}E(t){\rangle}$ computed in the absence of noise. To the
extent that the former quantity is larger, one can infer that noise enhances
the energy diffusion associated with periodic driving in the sense that the
mean change in energy is larger than what would obtain allowing only for the
separate effects of driving and noise.

The results of such an analysis are exhibited in Fig.~8, each panel of which
plots the quantity
\begin{equation} \label{deltaV}
{\Delta}V=\int {\langle}{\partial}V/{\partial}t{\rangle} ~dt,
\end{equation}
derived for the same set of initial conditions, evolved in the same potential
with the same amplitude $m_{0}$ and same ${\omega}_{0}$ but in the presence
of additive white noise with different amplitudes ${\eta}$. For both the
Plummer and Dehnen potentials, it is evident that, at least for low amplitudes
$m_{0}$, increasing ${\eta}$ tends to increase ${\Delta}V$, {\em i.e.,} that
the noise and driving reinforce one another in a superadditive fashion.
However, for higher amplitudes $m_{0}$, more variability is observed, the one
uniform feature being that the relative degree of reinforcement is less.
In this sense, one can say that {\it noise is more important as a source 
of energy diffusion in systems subjected to relatively low amplitude driving}.
\begin{figure}
\includegraphics[width=3in]{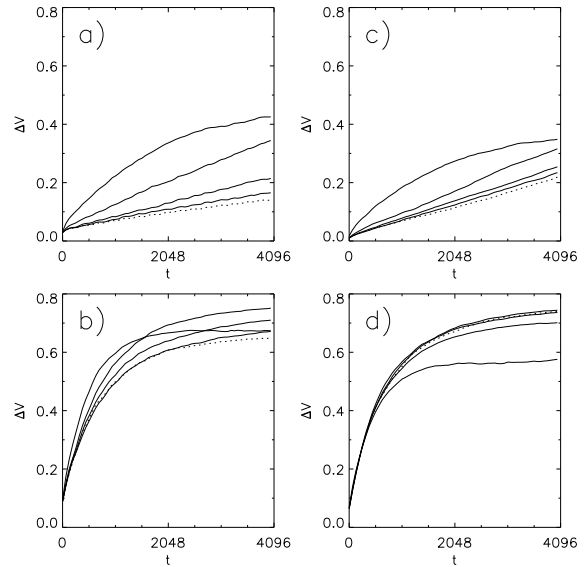} %fig8
\caption{Mean value of the quantity ${\Delta}V$, as defined in \ref{deltaV},
computed for $1200$ orbits with initial energy $E=-0.8$ evolved in a pulsating
triaxial Plummer potential with axis ratio $a^{2}:b^{2}:c^{2}= 1.25:1:0.75$,
${\omega}_{0}=1.3$, and variable amplitude (a) $m_0=0.1$ and (b) $m_0=0.2$.
Dashed lines correspond to integrations without white noise, while the
magnitude of white noise $\eta$ increases with increasing line thickness
through values $\eta=10^{-7}, 10^{-6}, 10^{-5}, 10^{-4}$.  (c) - (d) The 
same for a triaxial Dehnen potential with axis ratio $1.25:1:0.75$, and 
${\omega}_{0}=2.75$.
}
\end{figure}

But how is all this effected by allowing for colored noise with a finite
autocorrelation time?  If the autocorrelation time $t_{c}$ is sufficiently 
short compared with the dynamical time, colored noise has virtually the same 
effect as white noise.  However, when $t_{c}$ becomes comparable to the 
orbital time $t_{D}$ the effects of the noise become weaker; and, for $t_{c}$ 
sufficiently large compared with $t_{D}$ the noise has only a relatively 
minimal effect.

Figure 9 illustrates how, for the same initial conditions used to generate
Fig.~6, the relative number $f_{c}$ of chaotic orbits depends on both amplitude
${\eta}$ and autocorrelation time $t_{c}$. The top panel illustrates clearly
that, for fixed ${\eta}$, increasing $t_{c}$ reduces the degree to which the
noise converts regular orbits to chaotic. However, it is also apparent that,
for sufficiently large ${\eta}$, almost all the orbits still become chaotic.
The autocorrelation time must assume a value even larger than $t_{c}=32$ if
the effects of the noise are to `turn off' almost completely for amplitudes
as large as ${\eta}=10^{-4}$.
\begin{figure}
\includegraphics[width=3in]{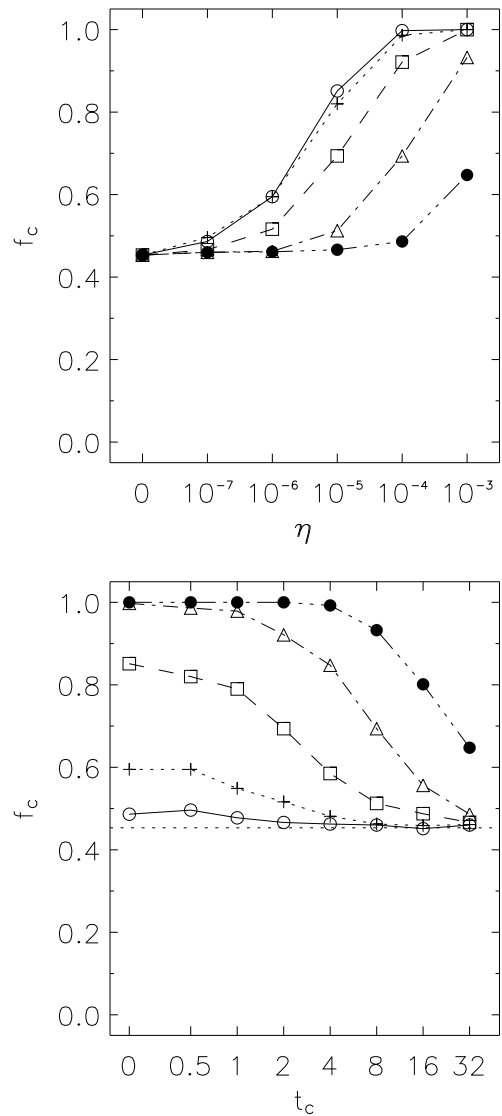} %fig9
\caption{
Relative measure of chaotic orbits, $f_{c}$, computed for a microcanonical
ensemble of $800$ initial conditions with energy $E=-0.8$ in a pulsating 
triaxial Plummer potential with axis ratio $a^{2}:b^{2}:c^{2}=1.25:1:0.75$, 
evolved with frequency ${\omega}_{0}=1.3$ and $m_{0}=0.1$ in the presence of 
noise with varying amplitude ${\eta}$ and autocorrelation time $t_{c}$.  
In (a): solid line with empty circles represents white noise 
(limit $t_c \to 0$), short-dashed line with crosses $t_c=0.5$, long-dashed 
line with squares $t_c=2$, and dash-dot line with triangles $t_c=8$, 
dash-dot-dot-dot line with filled circles $t_c=32$.  In (b): solid line with 
empty circles represents $\eta=10^{-7}$, short-dashed line with crosses 
$\eta=10^{-6}$, long-dashed line with squares $\eta=10^{-5}$, dash-dot line 
with triangles $\eta=10^{-4}$, dash-dot-dot-dot line with filled circles 
$\eta=10^{-3}$.  Horizontal dashed line corresponds to the fraction of chaotic 
orbits for the same ensemble evolved in the absence of any noise.
}
\end{figure}

The panel a) of Fig.~10, generated for orbits in triaxial Plummer, demonstrate 
that the rates of transition between sticky and wildly chaotic behavior are 
largely independent of the autocorrelation time $t_c$.  The panel c) of 
Fig.~10, representing orbits in triaxial Dehnen potentials, show that 
increasing the autocorrelation time $t_{c}$ tends also to weaken the effects 
of noise in accelerating transitions between sticky and wildly chaotic behavior.
\begin{figure}
\includegraphics[width=3in]{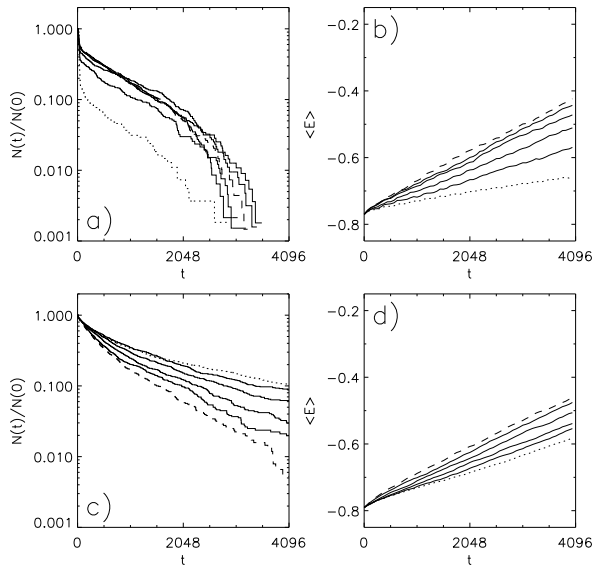} %fig10
\caption{
(a) Fraction $N(t)/N(0)$ of chaotic orbits that remain `sticky' after time $t$,
computed for ensembles of $800$ initial conditions with $E=-0.8$, evolved in a 
triaxial Plummer potential with axis ratio $a^{2}:b^{2}:c^{2}=1.25:1:0.75$, 
${\omega}_{0}=1.3$, and $m_{0}=0.1$, allowing for colored noise with 
${\eta}=10^{-4}$ and variable $t_{c}$.  The magnitude of $t_{c}$ increases 
with increasing line thickness through values $t_{c}=0.5$, $1$, $2$, and $4$. 
The long-dashed line corresponds to white noise. The short-dashed line 
corresponds to the same ensemble evolved in the absence of any kind of noise. 
(b) The mean energy ${\langle}E(t){\rangle}$ computed for the same ensembles.
(c) and (d) The same for ensembles evolved in a triaxial Dehnen potential
with $a^{2}:b^{2}:c^{2}=1.25:1:0.75$ and ${\omega}_{0}=2.75$, again with
$m_{0}=0.1$ and ${\eta}=10^{-5}$.
}
\end{figure}

The panels b) and d) of Fig.~10, generated for the same initial conditions,
demonstrates that increasing autocorrelation time $t_c$ decreases the rates 
of the mean change in energy.  For both potentials, the curve for $t_{c}=0.5$ 
coincides very nearly with the white noise curve, and the appreciable 
decreases in ${\langle}E(t){\rangle}$ arise for $t_{c}{\;}{\ge}{\;}1.0$.

\section{FREQUENCY VARIATIONS}
The objective now  is to explore how the results described in the preceding
Section are altered if one allows instead for variations in the pulsation
frequency, again modeled as Gaussian noise sampling an Ornstein-Uhlenbeck
process. What this entails is allowing for a variable frequency
${\omega}(t)={\omega}_{0}+{\delta}{\omega}(t)$, where the random frequency
shift ${\delta}{\omega}$ is characterized by moments
\begin{displaymath}
{\langle}{\delta}{\omega}(t){\rangle} = 0
\end{displaymath}
and
\begin{equation}
{\langle}{\delta}{\omega}(t_{1}){\delta}{\omega}(t_{2}){\rangle} =
{\Omega}^{2}\exp(-|t_{1}-t_{2}|/t_{c}).
\end{equation}
This implies in particular that ${\langle}{\omega}^{2}(t){\rangle}=
{\omega}_{0}^{2}+{\Omega}^{2}$.
The form of the experiments was the same as in the preceding Section.  The 
computations were performed assuming $F^{i}={\eta}=0$, {\it i.e} no ordinary 
noise.

As discussed in the preceding Section, allowing for `ordinary' white noise
tends to increase the relative abundance of chaotic orbits, especially for
systems which, in the absence of driving, admit few if any chaotic orbits.
Allowing for a `noisy' frequency with vanishing autocorrelation time has 
exactly the opposite effect. For systems where, in the absence of driving,
little if any chaos is present, making the frequency `stutter' by introducing
even comparatively low amplitude white noise impulses tends instead to
{\em decrease} the relative abundance of chaotic orbits. This is, {\em e.g.,}
evident from Fig.~11, which was generated for the same initial conditions
as Fig.~6, now allowing for white noise variations in ${\omega}$ with
different amplitudes ${\Omega^2}$.
\begin{figure}
\includegraphics[width=3in]{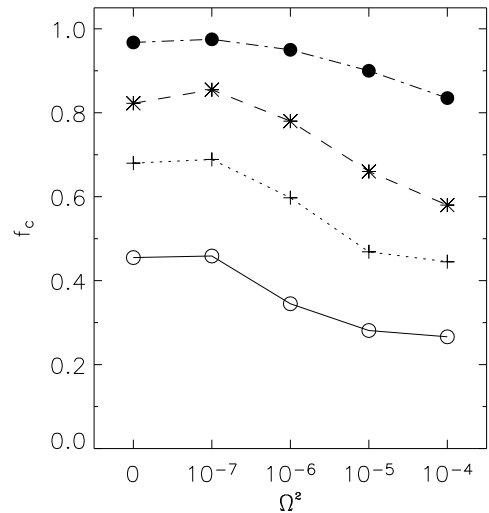}%fig11
\caption{
Relative measure of chaotic orbits, $f_{c}$, computed for a microcanonical
ensemble of $800$ initial conditions with energy $E=-0.8$ in a pulsating 
triaxial Plummer potential with axis ratio $a^{2}:b^{2}:c^{2}=1.25:1:0.75$, 
evolved with ${\omega}_{0}=1.3$, $m_{0}=0.1$ (empty circles with solid line),
$m_{0}=0.15$ (crosses with short-dashed line), $m_{0}=0.2$ (asterisks with
long-dashed line) and $m_{0}=0.3$ (filled circles with dash-dotted line),
allowing for `white noise' random variations in frequency with variable 
amplitude ${\Omega^2}$.
}
\end{figure}

As is illustrated in Fig.~12, allowing for random variations in frequency
tends also to suppress transitions from sticky to wildly chaotic behavior
and to decrease the amount of energy that the driving can pump into the
orbits.  White noise variations in frequency dramatically reduce the effects
of driving as a source of energy diffusion.
\begin{figure}
\includegraphics[width=3in]{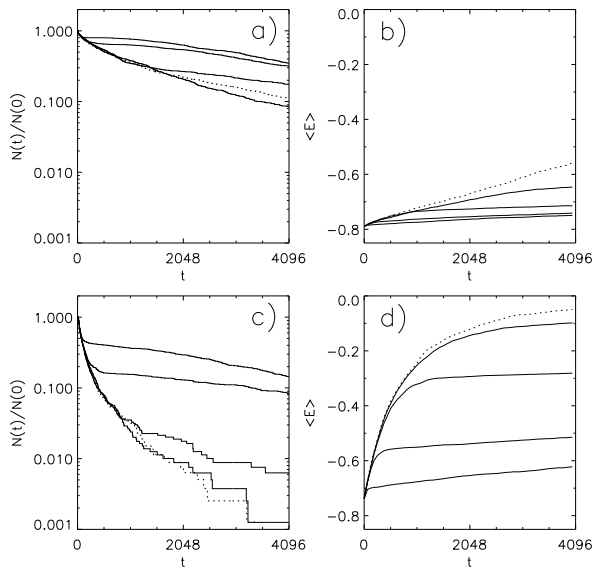}%fig12
\caption{
Fraction $N(t)/N(0)$ of chaotic orbits that remain `sticky' after time $t$,
computed for ensembles of $800$ initial conditions with $E=-0.8$, evolved in
a triaxial Dehnen potential with axis ratio $a^{2}:b^{2}:c^{2}=1.25:1:0.75$,
${\omega}_{0}=2.75$, and $m_{0}=0.1$, allowing for `white noise' variations
in frequency with variable amplitude ${\Omega}$. The magnitude of the noise
increases with increasing line thickness through values ${\Omega^2}=10^{-7}$,
$10^{-6}$, $10^{-5}$, and $10^{-4}$. The dashed line corresponds to
integrations without frequency variations. (b) The mean energy
${\langle}E(t){\rangle}$ computed for the same ensembles. (c) and (d) The
same for $m_{0}=0.2$.
}
\end{figure}

These differences are reflected by the fact that even very weak
perturbations, with amplitudes as small as ${\Omega^2}=10^{-6}$, completely
alter the Fourier spectra of the orbits. Examples thereof are provided
in Fig.~13 for both triaxial Plummer and triaxial Dehnen potentials. In
each case, it is evident that, in the absence of the perturbation, there is a
dominant resonance, ${\Omega}^{(1)}=\omega_0/2=0.65$ for the Plummer potential
and ${\Omega}^{(1)}=\omega_0/2=1.375$ for the Dehnen potential. But, in each
case, it is also apparent that, even for noise as weak as ${\Omega^2}=10^{-5}$,
the effects of this resonance have entirely disappeared.
\begin{figure}
\includegraphics[width=3in]{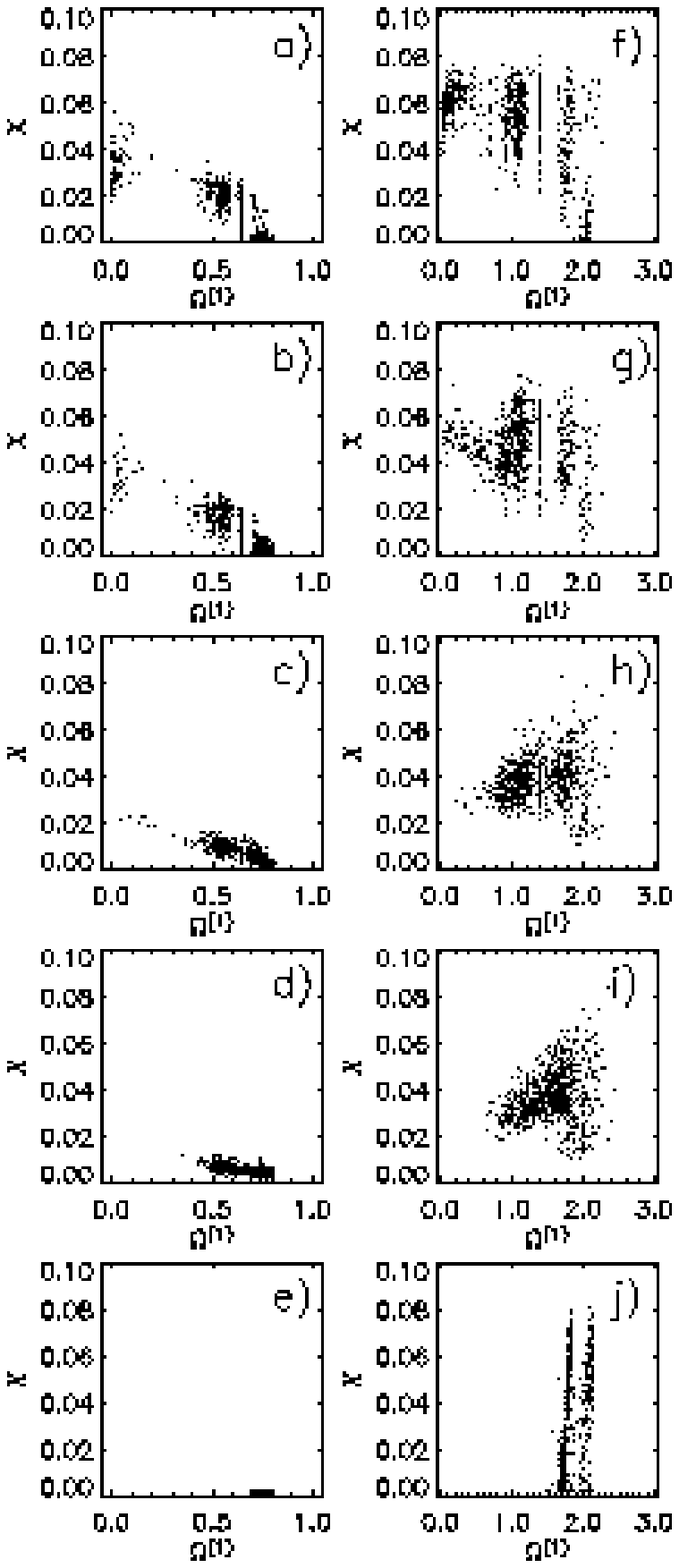}%fig13
\caption{
Finite time Lyapunov exponent as a function of frequency $\Omega^{(1)}$ for
$800$ orbits with initial energy $E=-0.8$ in a triaxial Plummer potential with
axis ratio $a^{2}:b^{2}:c^{2}=1.25:1:0.75$, pulsating with amplitude $m_0=0.1$ 
and $\omega_{0}=1.3$, allowing for `white noise' variations with (a) 
${\Omega}=0$, (b) ${\Omega^2}=10^{-7}$, (c) ${\Omega}=10^{-6}$, and 
(d) ${\Omega}=10^{-5}$. (e) The same ensemble evolved in the absence of 
both noise and driving.  (f) - (j) The same for orbits in a Dehnen potential 
with ${\omega}_{0}=2.75$, again with $a^{2}:b^{2}:c^{2}=1.25:1:0.75$ and
$m_{0}=0.1$.
}
\end{figure}

Also evident from this Figure is the fact that even a very small ${\Omega^2}$
suffices to prevent orbits from diffusing to very large energy. In the absence
of frequency variations, both potentials admit orbits that diffuse towards
very large radii and, consequently, have very small peak frequencies.
However, for ${\Omega^2}>10^{-5}$, there are no orbits with 
${\Omega}^{(1)}<0.4$. This does not, however, necessarily imply that
there are no chaotic orbits at all. For the case of the Dehnen potential
there are still significant numbers of chaotic orbits with larger values 
of ${\Omega}^{(1)}$.

One might naively have expected that the stuttering simply `turns off' the
effects of the driving, so that orbits pulsed with $m_{0}=0.1$ and
${\Omega^2}=10^{-7}$, $10^{-6}$, or $10^{-5}$ behave very much like orbits
evolved with $m_{0}={\Omega^2}=0$. This, however, is {\em not} the case.
It is, for example, clear that, for the Dehnen model, the distribution of
peak frequencies for an unpulsed, constant frequency integration has far 
more structure than the distributions for the pulsed, variable frequency 
models. Even though frequency variations can dramatically reduce the efficacy 
of the driving as a source of energy diffusion, they do not transform orbits 
back to the form they had in the absence of driving.

Adding color strengthens the effects of the perturbation. An example
of this tendency is illustrated in Fig.~14, which exhibits the fraction of
chaotic orbits for the ensembles of $800$ orbits in a pulsed Plummer
potential, allowing for random variations in frequency with different values
of ${\Omega^2}$ and $t_{c}$.  It is clear from the bottom panel that increasing
the value of autocorrelation time from the $t_{c}\to 0$ white noise limit
results in a systematic increase in the relative measure of chaotic orbits.
Also evident is the fact that even a relatively small value $t_{c}=0.5$ is
sufficient to significantly increase the number of chaotic orbits relative
to the white noise limit. As is evident from the lower panel, in the long
autocorrelation time limit, $t_c \to \infty$, the effects of the colored
noise gradually turn off.  This is to be expected, if one views
$t_c \to \infty$ as an adiabatic limit.  The time-scale on which this
turning off occurs is directly proportional to the amplitude of colored noise,
{\it i.e}, it takes longer autocorrelation times for the effects of the 
stronger noise to reach the adiabatic limit.
\begin{figure}
\includegraphics[width=3in]{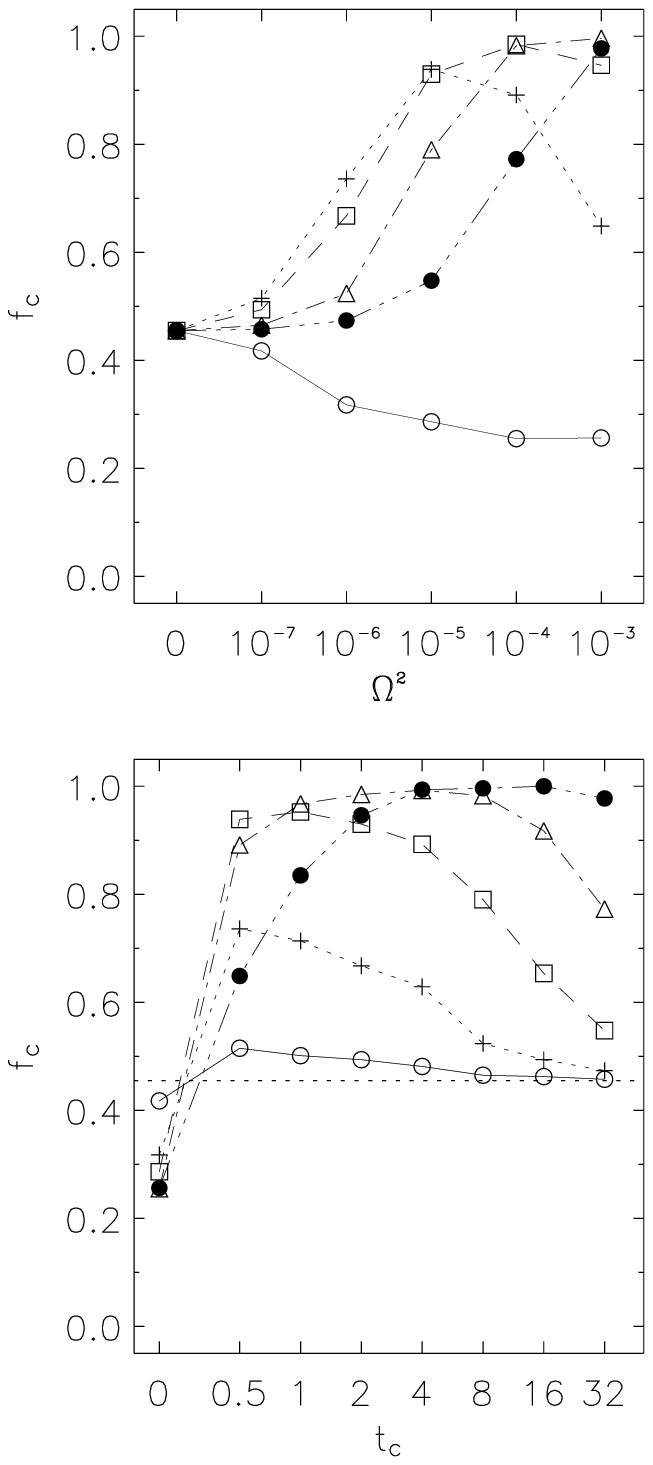}%fig14
\caption{
Relative measure of chaotic orbits, $f_{c}$, computed for a microcanonical
ensemble of $800$ initial conditions with energy $E=-0.8$ in a pulsating 
triaxial Plummer potential with axis ratio $a^{2}:b^{2}:c^{2}=1.25:1:0.75$, 
evolved with frequency ${\omega}_{0}=1.3$ and $m_{0}=0.1$, allowing for 
`colored noise' variations in frequency with variable amplitude ${\Omega^2}$ 
and autocorrelation time $t_{c}$.  In (a): solid line with empty circles 
represents white noise (limit $t_c \to 0$), short-dashed line with crosses 
$t_c=0.5$, long-dashed line with squares $t_c=2$, and dash-dot line with 
triangles $t_c=8$, dash-dot-dot-dot line with filled circles $t_c=32$.  
In (b): solid line with empty circles represents $\Omega^2=10^{-7}$, 
short-dashed line with crosses $\Omega^2=10^{-6}$, long-dashed line with 
squares $\Omega^2=10^{-5}$, dash-dot line with triangles $\Omega^2=10^{-4}$, 
dash-dot-dot-dot line with filled circles $\Omega^2=10^{-3}$.  Horizontal 
dashed line corresponds to the fraction of chaotic orbits for the same 
ensemble evolved in the absence of any noise.
}
\end{figure}

The left panel of Fig.~15 illustrates the fact that allowing for a finite
autocorrelation time enhances the transitions from sticky to wildly chaotic
behavior. The right hand panel illustrates the fact that, for fixed amplitude 
${\Omega^2}$, integrations with nonvanishing autocorrelation time $t_{c}$ 
will pump more energy into the orbits than corresponding integrations with
white noise or in the absence of noise.  The observed approach of the thick 
solid lines representing large autocorrelation time $t_c$ to the short-dashed 
lines corresponding to integrations in the absence of any noise illustrates 
that the effects of the noise are turning off in the $t_c \to \infty$ limit.
\begin{figure}
\includegraphics[width=3in]{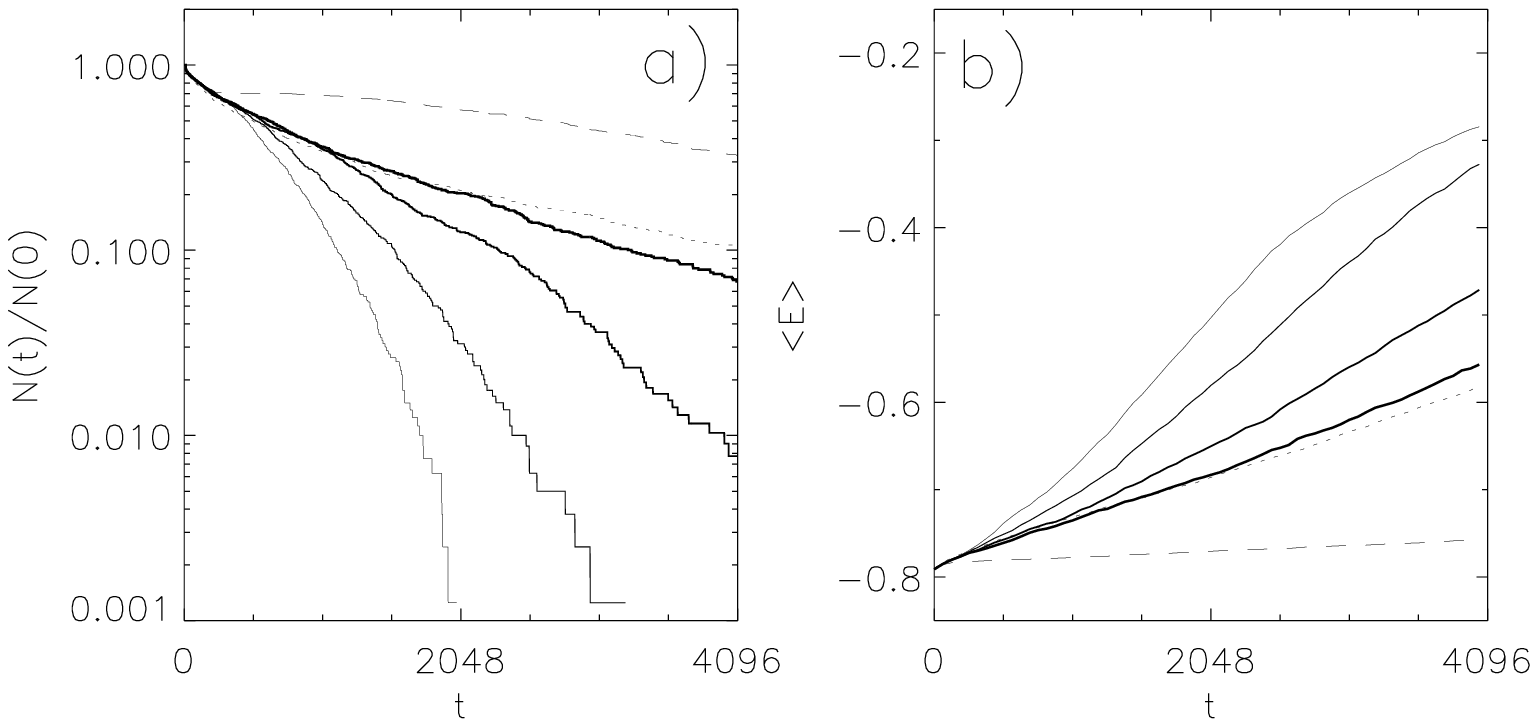}%fig15
\caption{
(a) Fraction $N(t)/N(0)$ of chaotic orbits in a $800$ orbit ensemble that
remain `sticky' after time $t$, computed for an ensemble of initial conditions 
with $E=-0.8$, evolved in a pulsating triaxial Dehnen potential with axis 
ratio $a^{2}:b^{2}:c^{2}=1.25:1:0.75$, ${\omega}_{0}=2.75$ and $m_{0}=0.1$, 
allowing for `colored noise' random variations in frequency with
${\Omega^2}=10^{-5}$ and variable $t_{c}$. The magnitude of $t_{c}$ increases
with increasing line thickness through values $t_{c}=0.5$, $2$, $8$, and $32$. 
The long-dashed line corresponds to white noise. The short-dashed line 
corresponds to the same ensemble evolved in the absence of any kind of noise.  
(b) The mean energy ${\langle}E(t){\rangle}$ computed for the same ensembles.  
}
\label{landfig}
\end{figure}

One other point would seem significant. For the case of `ordinary' noise,
incorporated as a random force in the equations of motion, typically a
comparatively large autocorrelation time $t_{c}{\;}{\approx}{\;}1$ is required
before one sees appreciable differences from the white noise limit. By
contrast, however, for the case of `noisy' variations in frequency even an
autocorrelation time as small as $t_{c}=0.5$ yields substantial differences
from the white noise limit.

\section{CONCLUSIONS}
Orbits in three-dimensional potentials subjected to periodic driving,
$V(x^{i},t)=(1+m_{0}\sin {\omega}t)V_{0}(x^{i})$, divide into two distinct
types, between which transitions are seemingly impossible: regular orbits, 
which are strictly periodic and have vanishing Lyapunov exponents, and chaotic 
orbits, which are aperiodic and have at least one positive Lyapunov exponent. 
Viewed over finite time intervals, the chaotic orbits divide in turn into 
two types: `sticky' chaotic segments, which are locked to the driving 
frequency and exhibit little systematic energy diffusion, and `wildly chaotic' 
segments, which are not so locked and can exhibit significant energy 
diffusion. The latter distinction is not absolute, transitions between sticky 
and wildly chaotic apparently involving motion through entropy barriers.

The relative abundance of sticky versus wildly chaotic segments depends
considerably on the form of the potential. However, sticky segments tend
overall to be more common for low amplitude driving, {\em i.e.,} the relative
phase space measure associated with sticky orbits tends to be larger.
Increasing the amplitude also tends to increase the transition rate between
sticky and wildly chaotic behavior, which suggests that the `holes'
associated with the entropy barriers become larger as $m_{0}$ increases.

The properties of orbits in such time-periodic potentials can be impacted
significantly even by very low amplitude perturbations, both random impulses
added to the equations of motion and random variations in the driving
frequency, in each case modeled as Gaussian noise sampling an
Ornstein-Uhlenbeck process.  For fixed amplitude of the perturbation, the
largest effects arise in the white noise limit of vanishing autocorrelation
time $t_{c}$ and decrease systematically with increasing $t_{c}$. The details
of the perturbation tend to be comparatively unimportant. For example,
for the case of `ordinary' noise the presence or absence of friction of
comparable magnitude and/or making the noise multiplicative as opposed to
additive has a comparatively minimal effect.  Even the precise strength of
the perturbation is relatively unimportant, the response exhibiting only a
weak, roughly logarithmic dependence on amplitude.

Ordinary noise tends to enhance the effects of periodic driving by increasing 
the relative measure of chaotic orbits, especially in potentials which contain 
little if any chaos in the absence of driving. It also acts to accelerate 
transitions between sticky and wildly chaotic behavior, thus facilitating 
more efficient energy diffusion.  The effects of noise in frequencies of 
pulsations have more complex effects.  White noise suppresses the effects of 
periodic driving by decreasing transitions between sticky and wildly chaotic 
behavior.  Colored noise fluctuations, at least for autocorrelation times on 
the order of the dynamical time, enhances the effects of periodic driving.  
These effects gradually turn off in the adiabatic $t_c \to \infty$ limit.

Ordinary noise appears to act by enhancing the effects of the resonances
associated with the periodic driving. Wiggling orbital trajectories
can allow an otherwise regular orbit to be impacted by a resonance
which it might not otherwise `feel', thus causing it to become chaotic.
And similarly, the accelerated rate of transitions between sticky and wildly
chaotic behavior can be attributed to the fact that such wiggling helps
already chaotic orbits `find' holes in the entropy barrier. Indeed, the effect
of noise in enhancing such transitions is very similar qualitatively to
the effect of noise as a source of accelerated diffusion through cantori in
time-independent two-degree-of-freedom Hamiltonian systems~\cite{pk99}.

Random white noise variations in the driving frequency appear to weaken the 
resonances. Making the frequency `stutter' by adding even a very low 
amplitude, high frequency component to the driving significantly alters
the Fourier spectra of orbits, suppressing structure which, in the absence
of the perturbation, indicates the importance of the resonances. Variations
characterized by non-vanishing autocorrelation time enhances the resonances
because the Fourier spectrum of the frequency now has another strong component
with which the fundamental frequencies of orbits can resonate.

\begin{acknowledgments}
Partial financial support was provided by NSF AST-0307351.  We would like to 
thank Florida State University School of Computational Science and Information 
Technology for granting access to their supercomputer facilities.
\end{acknowledgments}

\vfill

\appendix
\section{Implementing White and Colored Noise}

Here we outline the implementation of noisy equations of motion 
(\ref{langevin}).  The term $F^i(t)$ is a Gaussian-distributed random force 
which samples Ornstein-Uhlenbeck process.  It has a autocorrelation time 
$t_c$, which in the limit $t_c \to 0$ samples ordinary white noise.  
$F^i(t)$ is a normalized solution to the free particle Langevin equation
\begin{equation} \label{lan2}
{{d u^i}\over{d t}} + {1 \over t_c} u^i = {1 \over t_c} w^i(t),
\end{equation}
where $w^i(t)$ is Gaussian-distributed white noise with zero mean and unit 
variance.  It is well known that such a variable is easily generated as a 
sum of 12 random numbers in the range $[-1/2,1/2]$.  The solution to 
(\ref{lan2}) is given by \cite{uo}
\begin{eqnarray}
u^i(t)  = & u^i_0 &~ \exp(-t/t_c) \nonumber \\
    & + & {1\over t_c}~ \exp(-t/t_c)\int\limits_0^t
\exp(\xi/t_c)~ w^i(\xi)~ d\xi,
\end{eqnarray}
which, after some straightforward algebra, yields
\begin{equation} \label{norm}
\langle {{u^i}^2} \rangle = {1\over {2 t_c}} +\left({{u_0}^i}^2-{1\over
{2 t_c}}\right) \exp(-2t/t_c).
\end{equation}
The second term is transient, so that for $t \gg t_c$,
\begin{equation} \label{norm2}
\langle {{u^i}^2}\rangle \approx {1\over {2 t_c}}.
\end{equation}
This means that the random variable $\sqrt{2 t_c}{~\!} u^i$ has 
autocorrelation time $t_c$, zero mean and unit variance. 

We determine the proper amplitude of colored noise $A$ after substituting 
$F^i(t)=\sqrt{A} {~\!} u^i(t)$, into the equation (\ref{Fc}) and equating 
it with the equation (\ref{Kc}):
\begin{eqnarray} \label{noisy}
\langle F^i(t_1) F^j(t_2) \rangle & = & 
{A\over{2 t_c}} \exp(-|t_1-t_2|/t_c) \nonumber \\
& = & \Delta^2 \exp(-|t_1-t_2|/t_c) \nonumber \\
& = & {D\over{2 t_c}} \exp(-|t_1-t_2|/t_c),  
\end{eqnarray}
from which it is clear that $A=D\equiv 2 \Theta \eta$.  Therefore, the 
properly scaled noise noise factors are 
$F^i(t) = \sqrt{2 \Theta \eta}{~\!} u^i(t)$, where $u^i(t)$ is the 
solution of (\ref{lan2}) for colored noise, and 
$F^i(t) = \sqrt{2 \Theta \eta}{~\!} w^i(t)$, 
where $w^i(t)$ is Gaussian-distributed white noise with zero mean and unit 
variance.


\begin{references}

\bibitem{LB}D.~Lynden-Bell, Mon.~Not.~R.~Astron.~Soc., {\bf 136}, 101 (1967).
\bibitem{bertin}G.~Bertin, {\em Dynamics of Galaxies} (Cambridge University
Press, Cambridge, 2000).
\bibitem{reiser}M.~Reiser, {\em Theory and Design of Charged Particle Beams}
(Wiley, New York, 1994).
\bibitem{gluckstern}R.~L.~Gluckstern, Phys. Rev. Lett. {\bf 73}, 1247 (1994).
\bibitem{kvs}H.~E.~Kandrup, I.~M.~Vass, and I.~V.~Sideris,
Mon.~Not.~R.~Astron.~Soc. {\bf 341}, 927 (2003).
\bibitem{tk}B.~Terzi{\'c} and H.~E.~Kandrup, Mon.~Not.~R.~Astron.~Soc.,
in press (2003). (astro-ph/0306323)
\bibitem{mmp}R.~S.~MacKay, J~D.~Meiss, I.~C.~Percival, Physica {\bf 13D}, 1
(1984).
\bibitem{lw}A.~J.~Lichtenberg and B.~P. Wood, Phys.~Rev.~Lett. {\bf 62},
2213 (1989).
\bibitem{pk99}I.~V.~Pogorelov and H.~E.~Kandrup, Phys. Rev. {\bf E 60},
1567 (1999).
\bibitem{BS}C.~L.~Bohn and I.~V. Sideris, Phys.~Rev.~Lett., submitted (2003).
\bibitem{vanK}N.~G.~van Kampen, {\em Stochastic Processes in Physics and
Chemistry} (North Holland, Amsterdam, 1981).
\bibitem{deh}W.~Dehnen, Mon.~Not.~R.~Astron.~Soc. {\bf 265}, 250 (1993).
\bibitem{lauer}T.~Lauer {\em et al}, Astron.~J. {\bf 110}, 2622 (1995).
\bibitem{bgs}G.~Bennettin, L.~Galgani, and J.~M.~Strelcyn, Mecannica
{\bf 15}, 9  (1980).
\bibitem{inf}Unless the orbit becomes energetically unbounded and escapes
towards infinity, where $t_{D}\to\infty$.
\bibitem{weird}For very weak amplitude driving, most of the wildly chaotic
orbits do not
immediately diffuse to very large energies and radii; and, in this case,
selecting a `sticky' energy range that is too large misclassifies many wildly
chaotic orbits as sticky. Hence the erroneous prediction of an anomalously
small transition rate.
\bibitem{haber}I.~Haber, D.~Kehne, M.~Reiser, and H.~Rudd, Phys. Rev.
{\bf A 44}, 5194 (1991).
\bibitem{sk02}I.~V.~Sideris and H.~E.~Kandrup, Phys. Rev. {\bf E 65}, 066203
(2002).
\bibitem{uo} G.~E.~Uhlenbeck and L.~S.~Ornstein, Phys. Rev. {\bf 36}, 823 
\end{references}
\end{document}